\documentstyle[preprint,aps]{revtex}
\begin{document}
\draft
\preprint{PRL in press}
\title{Probing Spatial Correlations with Nanoscale Two-Contact Tunneling}
\author{Jeff M. Byers}
\address{Department of Physics, University of California, Santa Barbara,
         California 93106}
\author{Michael E. Flatt\'{e}}
\address{Division of Applied Sciences, Harvard University, Cambridge,
         Massachusetts 02138}
\date{February 22, 1994}
\maketitle
\begin{abstract}
Interference effects on the transport through two localized
tunnel junctions on the surface of a well-grounded sample reveal
intrinsic spatial correlations characteristic of the uncoupled sample.
Differential conductances of the two-junction probe are
related to the spatial correlations of
both normal and superconducting samples.
For a superconducting sample the gap anisotropy strongly affects the results.
This may serve as a sensitive probe of the order parameter in
high-temperature superconductors.
\end{abstract}
\narrowtext

The advent of scanning tunneling microscopy (STM)
has enabled the characterization
of materials on the atomic scale through
measurements of the local density of states (LDOS).
Recently, in a series of STM experiments on a Cu (111) surface,
the local electronic correlations of the surface-state
electrons were probed through their influence on the LDOS
around an Fe impurity\cite{Crommie1}.
Properties which might be determined from these types of
measurements, but could not be probed by an STM
measurement on the homogeneous sample,
include the {\it angularly-resolved}
dispersion relations and mean free path, as well as the  density of
states as a function of energy {\it and} momentum.

In this type of STM experiment
the characteristics of an impurity limit
the information available on the local electronic
correlations around it.
In particular, the influence of the impurity is manifested
in oscillations (of approximately
the Fermi wavelength) in the LDOS.
These short-distance oscillations allow accurate determination
of the dispersion relations, but hinder measurements of
long-distance properties.
By replacing the impurity with a contact, and taking an
appropriate differential conductance between the two
contacts, one obtains
a measurement which does not contain substantial oscillations, but
reveals long-distance behavior.

A conceptually straightforward, but impractical, two-contact
arrangement would consist of two STM tips
which could be placed from $1$\AA\ to $1000$\AA\ apart. This
would allow probing of correlations with short length scales
($1-10$\AA),
such as Fermi wavelengths in desired
directions, medium length scales ($10-100$\AA), such
as high-$T_c$ superconducting coherence lengths and
charge-density wave oscillations, and
long length scales ($100-1000$\AA) such as mean free
paths, transitions from ballistic to diffusive propagation,
low-$T_c$ coherence lengths, charge-density-wave correlation
lengths, and angularly anisotropic density-of-states
\hbox{effects\cite{Byers}.}

Practical alternatives to the two-tip STM are
possible. If a single small contact of size $\sim{100}$\AA\
could be made on a surface, the other contact could be an
STM. In some systems contact has been made to nanofabricated wires
as small as $100$\AA, fabricated using STM-CVD\cite{STMCVD}
on a smooth surface.
With such an arrangement at least the long-range correlations could
be probed.

The two-contact experiment is also insensitive to a
small concentration of impurities on the surface. Impurities
would produce oscillations in the LDOS due to scattering of
electrons from one contact to the other. A contact detects
an average of the LDOS over its area, so if the contact
diameter is large compared to the Fermi wavelength these
oscillations would be greatly suppressed.

The specific application we will focus on is detecting gap
anisotropy in high-temperature superconductors. For a
$d_{x^2-y^2}$ gap, which has four nodes, we find at voltages
much less than the gap quasiparticles can only travel in the
real-space directions roughly parallel to node momenta,
yielding ``channels'' of conductance. At voltages slightly
higher than the gap maximum there are more states for
momenta near the gap maximum, so the channels would appear
rotated by $45^o$. With a $100$\AA\ contact and an STM tip
$1000$\AA\ away, the angular resolution would be $6^o$,
similar to photoemission\cite{Shen}. Major advantages over
photoemission include the improved energy resolution and the
ability to characterize the surface with the STM while
performing the experiment. In contrast to other tunneling
probes of gap anisotropy\cite{Wollman}, the details of the
tunneling barrier are not important.

Measurements of gap anisotropy promise to be
effective in distinguishing among the various theories of
high-temperature superconductivity, including
phonon-mediated $s$-wave\cite{Eliashberg},
antiferromagnetic-spin-fluctuation-mediated
$d_{x^2-y^2}$\cite{SP}, or anyonic $d_{x^2-y^2}+i\alpha d_{xy}$\cite{anyon}.
Despite evidence of a finite
density of states at low energy\cite{Hardy}, angular
anisotropy of the gap\cite{Shen}, and changes of sign in the
gap around the Fermi surface\cite{Wollman}, evidence for
$d$-wave superconductivity is not conclusive. Furthermore,
even less evidence selects among the various forms
of $d$-wave gaps.

Figure 1 shows the geometry of the two-contact probe.
There are three reservoirs connected to
a sample.  One of the reservoirs is
well-connected to the sample and acts as a ground that determines its chemical
potential (this will be discussed more below).
This configuration is
intentionally different from the ungrounded, two-contact geometry
relevant for quantum dots\cite{Yigal1}.
In this Letter the behavior of the homogeneous sample material is of
interest - not the behavior of electrons confined to a small island.
The remaining two reservoirs are weakly linked to the sample
and act as tunnel junctions.

The Hamiltonian for the uncoupled system is
\begin{equation}
H_I\; = \; {\sum_{\alpha_i}}{\epsilon}_{\alpha_i}\:
c^{\dagger}_{\alpha_i}c_{\alpha_i}
    +  {\int}d{{\bf x}}d{\bf x}^{\prime}
    \Bigl[\,\sum_s{\epsilon}({\bf x},{\bf x}^{\prime})
     \psi_{s}^{\dagger}({\bf x}){\psi_{s}}({\bf x}^{\prime})
   + \Bigl( \Delta({{\bf x}},{{\bf x}}^{\prime})
\psi_{\uparrow}^{\dagger}({\bf x})
     \psi_{\downarrow}^{\dagger}({{\bf x}}^{\prime}) + h.c.\Bigr)\Bigr],
\label{HI}
\end{equation}
where $c^{\dagger}_{\alpha_i}$
creates an electron
in eigenstate $\alpha_i$ in lead $i=1$, $2$,
or $G$ (Ground).
The last terms describe the sample, and are
written in coordinate representation
in anticipation of a spatially inhomogeneous
response to the localized tunneling probes.  The transfer Hamiltonian
is
\begin{equation}
H_T  =  {\sum_{\alpha_i}}{\int}d{\bf x}[W_{\alpha_i}\upsilon_i({\bf x}
-{\bf x_i})
      {{\psi}^{\dagger}}({\bf x}){c_{\alpha_i}} +
{\rm h.c.}]
\label{HT}
\end{equation}
where $W_{\alpha_i}\upsilon_i({\bf x}-{\bf x_i})$
is the amplitude for an electron in energy level
$\alpha_i$ of lead $i$ to jump to position
${\bf x}$ in the sample.
${\bf x}_i$ is the mean position of
junction $i$. Since the ground is visualized as a broad
contact, we assume for simplicity $\upsilon_G({\bf x}) = \upsilon_G$.
The functions $\upsilon_1({\bf x})$ and
$\upsilon_2({\bf x})$ represent the geometry of the
junction interfaces (acting to an extent as the wave function of an electron
under the lead).

Each of the two localized junctions $1$ and $2$
has its own voltage difference with respect to ground ($V_1$
and $V_2$), as depicted
in Fig.\ \ref{fig1}.
The current through the junctions was calculated as a
function of voltage and position.
\begin{eqnarray}
I_1 & = & {4e\over{ \pi \hbar }} \int d\omega
\biggl[ {\mbox{f}_1}^+(\omega) \Bigl(
\Gamma_1 Im(g^{ret}({\bf x}_1,{\bf x}_1)) +
2\Gamma_1^2\Big[Im(g^{ret}({\bf x}_1,{\bf x}_1))\Big]^2  \Bigr)
\nonumber\\
&   & + \Bigl( {\mbox{f}_1}^+(\omega) + {\mbox{f}_2}^+(\omega) \Bigr)
\Gamma_1\Gamma_2\Big[Im(g^{ret}({\bf x}_1,{\bf x}_2))\Big]^2
  - \Bigl({\mbox{f}_1}^+(\omega) - {\mbox{f}_2}^+(\omega) \Bigr)
\Gamma_1\Gamma_2\Big[Re(g^{ret}({\bf x}_1,{\bf x}_2))\Big]^2
\nonumber\\
&   & + \Bigl( {\mbox{f}_1}^+(\omega) + {\mbox{f}_1}^-(\omega) \Bigr)
\;\:\Gamma_1^2\;\:\Big[Im(f^{ret}({\bf x}_1,{\bf x}_1))\Big]^2
  - \Bigl({\mbox{f}_1}^+(\omega) - {\mbox{f}_1}^-(\omega) \Bigr)
\;\:\Gamma_1^2\;\:\Big[Re(f^{ret}({\bf x}_1,{\bf x}_1))\Big]^2
\nonumber\\
&   & + \Bigl( {\mbox{f}_1}^+(\omega) + {\mbox{f}_2}^-(\omega) \Bigr)
\Gamma_1\Gamma_2\Big[Im(f^{ret}({\bf x}_1,{\bf x}_2))\Big]^2
  - \Bigl({\mbox{f}_1}^+(\omega) - {\mbox{f}_2}^-(\omega) \Bigr)
\Gamma_1\Gamma_2\Big[Re(f^{ret}({\bf x}_1,{\bf x}_2))\Big]^2
\biggr]
\label{thirdI1}
\end{eqnarray}
where
$g^{ret}({\bf x}_i,{\bf x}_j)={\int}d{\bf x}d{\bf x}^\prime
\upsilon_i({\bf x}-{\bf x_i})\bar g^{ret}({\bf x},{\bf x}^\prime)
\upsilon_j^* ({\bf x}^\prime-{\bf x_j})$, and $\bar g$ is the
Green's function of the uncoupled sample. $f^{ret}$ is
defined similarly.
${\mbox{f}_i}^{\pm} (\omega) = \mbox{f}(\omega \pm eV_i)
- \mbox{f}(\omega)$, where
$\mbox{f}(\omega)$ is the Fermi factor and
$\Gamma_i = \pi\sum_{\alpha_i} |W_{\alpha_i}|^2\delta(\omega-
\epsilon_{\alpha_i})$.
Each term in Eq.\ (\ref{thirdI1}) can be obtained from a Fermi's golden
rule calculation carried out to second order.
Equation\ (\ref{thirdI1}) represents several tunneling mechanisms
(depicted in Fig.\ \ref{fig2} (a) and (b)) that combine
wave function interference effects and transport.
The first and second terms
are direct electron tunneling (D) and reflection (R), respectively,
at junction 1.  These terms are not influenced by the second junction.

The second line of Eq.\ (\ref{thirdI1}) is the
influence of the other tunnel junction on the normal conduction channel (T).
If this line is rearranged, the combination
${\mbox{f}_1}^+(\omega)\Gamma_1\Gamma_2\big([Im(g^{ret})]^2
- [Re(g^{ret})]^2\big)$
represents an interference effect caused by the second junction as if it were
a block of material on the surface of the sample with no external connection
(this process is independent of $eV_2$).  The remaining combination
${\mbox{f}_2}^+(\omega)\Gamma_1\Gamma_2\big([Im(g^{ret})]^2
+ [Re(g^{ret})]^2\big)$
represents transport between junctions 1 and 2 through the sample and is
not present if $eV_2 = 0$.

The last two lines in Eq.\ (\ref{thirdI1}) can also
be rearranged into interference and transport terms in the same manner.
The third line of Eq.\ (\ref{thirdI1}) represents Andre\'{e}v reflection
\cite{Andreev} at junction 1 where
an incident electron in the lead is reflected as a hole with a Cooper
pair injected into the superconducting sample (AR)\cite{othAR}.
The last line in Eq.\ (\ref{thirdI1}) represents
a novel process where an incident electron at junction 1 is
Andre\'{e}v reflected but the hole ends up in the opposing lead (AT).
Alternatively, one can turn the reflected
hole in junction 2 into another incident electron and visualize the
process as each lead contributing an electron towards forming a
Cooper pair in the sample.
In order for this particular mechanism to contribute to the current
the two junctions must be within a few superconducting coherence lengths
of one another.

The Fermi's golden rule
calculations are justified if the sample has a definite
chemical potential. The grounding lead produces this
situation by inducing a lifetime $\Gamma_G^{-1} =
(2\pi |W_G\upsilon_G|^2N_G(0))^{-1}$ for an electronic excitation
to leave the system via the ground (where $N_i(0)$ is the
density of states at the Fermi level of lead $i$). In the
situation we consider $\Gamma_G$ is small enough not to
affect local electronic transport, but large enough that any
path from contact $1$ to contact $2$ which involves
scattering off the sample boundaries is strongly
suppressed.
The grounding-lead lifetime differs from those due to
intrinsic sample processes, such as inelastic scattering,
which produce different electronic excitations in the
sample. One needs only observe that in the ungrounded,
two-contact geometry, the current through lead $1$ must
equal the current through lead $2$, something not required
for the grounded geometry.

To remove the effect of the position-independent background a
cross-junction differential conductance can be defined by taking the
derivative of $I_1$ with respect to the voltage $V_2$ across the other
junction.  Doing so yields the simpler result
\begin{eqnarray}
R_Q{d\over{dV_2}}I_1({\bf x}_1,{\bf x}_2;V_2) = 4\Gamma_1 \Gamma_2
            \Big[ {|g^{ret}({\bf x}_1,{\bf x}_2;\omega = eV_2)|}^2
                - {|f^{ret}({\bf x}_1,{\bf x}_2;\omega = eV_2)|}^2
            \Big]
\label{result}
\end{eqnarray}
where $R_Q \equiv { {\pi \hbar}/{e^2}}$.
The relative sign beween the terms has its
physical origin in the Andre{\'{e}}v process (represented by ${|f^{ret}|}^2$)
which places a hole in the other lead, whereas
the normal channel (represented by ${|g^{ret}|}^2$)
places an electron there, thus resulting in opposing currents.
$I_2$ can be found by interchanging the labels 1 and 2 in $I_1$.
Forming the combinations $I_T=(I_1 - I_2)/2$
and $I_G=I_1+I_2$, different pieces of the Green's functions can be measured:
\begin{eqnarray}
 R_Q{d\over{dV_2}}I_T({\bf x}_1,{\bf x}_2;V_2) =
   \sigma_{BT} + 4\Gamma_1 \Gamma_2
   \Big[\big[Re\big(g^{ret}({\bf x}_1,{\bf x}_2;\omega
= eV_2)\big)\big]^2
   -\big[Im\big(f^{ret}({\bf x}_1,{\bf x}_2;\omega = eV_2)\big)\big]^2
 \Big]\nonumber\\
 {{R_Q}\over{2}}{d\over{dV_2}}I_G({\bf x}_1,{\bf x}_2;V_2) =
   \sigma_{BG} + 4\Gamma_1 \Gamma_2
   \Big[\big[Im\big(g^{ret}({\bf x}_1,{\bf x}_2;\omega
= eV_2)\big)\big]^2
   -\big[Re\big(f^{ret}({\bf x}_1,{\bf x}_2;\omega = eV_2)\big)\big]^2
 \Big]
\label{result2}
\end{eqnarray}
where $\sigma_{BT}$ and $\sigma_{BG}$ are position-independent background
terms.
Equations\ (\ref{result}) and\ (\ref{result2}) demonstrate that this
particular example of a nanofabricated probe
can make direct measurements of spatial correlations due to
electronic transport within the sample.

Figure\ \ref{fig2}(c) shows the distinction, for a
quasi-two-dimensional system,
between an isotropic superconductor's response
and the normal state response in the
cross-junction differential conductance. The bias is set
just above the gap maximum, $eV=1.1\Delta_0$.
For this Figure, to show the short-wavelength oscillatory
behavior, we took $\upsilon_1({\bf x}) = \upsilon_2({\bf
x}) = \delta({\bf x})$.
As the sample becomes superconducting the oscillations are enhanced
due to the increased density of states just above $\omega = \Delta_0$.
The wavelength of the rapid oscillation is set by $k_F$ and the asymptotic
behavior is $1/\rho$ as expected for a 2-D geometry. In a
typical high-$T_c$ superconductor, $k_F^{-1} \sim 1$\AA.
A characteristic estimate of the magnitude of this conductance
relative to the conductance through {\it one} of the contacts is $\Gamma
N(0)$, where $N(0)$ is the (normal) sample's density of
states.  This is of the order of a percent for most systems,
and should therefore be visible. Essentially, there is no
reason for the cross-junction conductance to be
substantially weaker than any other second-order tunneling
process. Many of these second-order processes, like
Andre\'ev processes within an ultrasmall junction\cite{Aproc}, have been
observed.

As a further application of Eq.\ (\ref{result}) consider the case of
two contacts on the surface
of a superconducting film that has a gap anisotropy, as the high-temperature
superconductors may have.
They are within an elastic
mean free path of one another so that the Green's functions represent ballistic
and not diffusive propagation.  At helium temperatures the
mean-free path on a high-temperature superconductor is
$10^3-10^4$\AA\cite{Klein}.

Figure\ \ref{fig3} shows the cross-junction differential
conductance
(multiplied by the radial distance $\rho$ to allow
easier visualization of the outlying regions)
for an anisotropic superconductor
as a function of $x$ and $y$ for the gap
$\Delta_{{\bf k}} = \Delta_0 \cos(2\phi_{\bf k})$.
$\phi_{\bf k}$ is the angle that
the momentum ${\bf k}$ makes with the crystallographic
{\it a}-axis.
The voltage bias is set well below the gap maximum ($eV=0.1\Delta_0$)
so that the
quasiparticles are only able to propagate in the directions where
$\Delta_{{\bf k}}$
has nodes.

One contact is $100$\AA\ in
diameter and the other is atomic scale (an STM tip).
To evaluate Eq.\ (\ref{result}) under these conditions
we take $\upsilon_1({\bf x}-{\bf x}_1)
= \theta( R-|{\bf x} - {{\bf x}}_1|)$ and
$\upsilon_2({\bf x}-{\bf x}_2)
= \delta ( {\bf x} - {{\bf x}}_2)$, where $R=50$\AA.
We calculated numerically
the Green's functions corresponding to a cylindrical Fermi surface
because a sample of high-temperature
superconductor would either be
a thin film or a layered structure with weakly-coupled planes.
The rapid oscillations
in the differential conductance on the scale of $k_F^{-1}$
are gone, but the long-distance correlations are still visible.
The dominant term from Eq.\ (\ref{result}) is
${|g^{ret}({\bf x}_1,{\bf x}_2;\omega = eV_2)|}^2$ for $eV_2\ll\Delta_0$.
The terms $Re(g^{ret})$ and $Im(g^{ret})$ are very similar in appearance,
therefore $I_G$ and $I_T$ look nearly identical to Fig.\ \ref{fig3}.

We note that for $eV > \Delta_0$ the enhanced density
of states at the anti-nodes causes channels rotated by $\pi/4$
from the sub-gap result.
These channels appear identical to those found in the impurity case
\cite{Byers}, with the important exception that the
short-wavelength oscillations present in
the impurity case are absent here. Also, the
signal in the two-contact experiment is substantially
greater.

The primary goal of this Letter has been to offer an example of how
two-contact localized spectroscopy can explore the spatial correlations
of an uncoupled sample.  Tunneling with an STM measures the local density of
states, however our results indicate the possibility of
probing {\it transport}
phenomena via STM imaging of a $100$\AA\ contact.
Measuring gap anisotropy is merely one possible application of the
ability to directly and in detail probe small-scale electronic transport.
Analysis of angularly-resolved mean free paths,
the transition between ballistic and diffusive
transport or the lifetimes of quasiparticles in both normal
and superconducting samples should also be possible with nanoscale
two-contact tunneling.

We acknowledge useful conversations with D. J. Scalapino and Y. Meir.
J. M. B. was supported by NSF Grant No.  DMR92-25027.
M. E. F acknowledges the support of JSEP through ONR N00014-89-J-1023.

{\it Note added in proof:} During revision we became aware
of similar work\cite{Niu}.

\begin{figure}
\caption{Probe-sample geometry.
The sample is strongly connected to ground, and weakly
connected to two localized junctions, at least one of which
is mobile. The possibility considered in this Letter
is a $100$\AA\ nanofabricated
contact for one localized junction (located at ${\bf x_1}$),
and an STM tip for the other (located at ${\bf x_2}$).}
\label{fig1}
\end{figure}
\begin{figure}
\caption{Physical processes occurring in two-junction
geometry.  (a) Schematic of normal state processes of direct
(D) and reflected (R)
tunneling between an electrode and the grounded sample. The process (T) is
the interference and transport effect between the electrodes.  (b) Schematic
of
anomalous processes in a superconducting sample where Andre\'{e}v reflection
occurs at one junction (AR) and both (AT).
(c)  Comparison of normal (solid line) and superconducting (dotted) state
in the cross-junction differential conductance (defined in text) at a bias
$eV = 1.1\Delta_0$.
The superconducting gap $\Delta=0.1\epsilon_F$ and is isotropic.
(Differential conductance is in units of
$4\Gamma_1 N(0) \Gamma_2 N(0){R_Q}^{-1}$).}
\label{fig2}
\end{figure}
\begin{figure}
\caption{Cross-junction differential conductance
in position space (units of ${k_F}^{-1}\sim 1$\AA)
multiplied by the radial distance $\rho$ from the origin. A
$100$\AA\ diameter contact is located at the origin (indicated
by the hole). The other contact is atomic-scale (an
STM tip). The function's value at a point indicates the
cross-junction differential conductance when the STM tip is
located there.
The sample is a superconductor with the
anisotropic gap $\Delta_{{\bf k}} = \Delta_0 \cos (2 \phi_{\bf
k})$
and $eV = 0.1\Delta_0$.
The conductance is in units of
$4\Gamma_1 N(0) \Gamma_2 N(0){(k_FR_Q)}^{-1}$.}
\label{fig3}
\end{figure}

\begin{references}
\bibitem{Crommie1} M.F. Crommie, C.P. Lutz and D.M. Eigler, Nature (London),
                   {\bf 363}, 524 (1993).
\bibitem{Byers}    J.M. Byers, M.E. Flatt\'{e} and D. J. Scalapino,
                   Phys. Rev. Lett. {\bf 71}, 3363 (1993).
\bibitem{STMCVD}   A.L. de Lozanne, E.E. Ehreichs and W.F.
		   Smith, J. Phys.: Condens. Matter {\bf 5}, A409
		   (1993); E.E. Ehrichs, W.F. Smith and A.L. de
Lozanne, Ultramicroscopy {\bf 42-44}, 1438 (1992).
\bibitem{Shen}     B.O. Wells, {\it et al.},
                   Phys. Rev. B {\bf 46}, 11830 (1992).
                   Z.-X. Shen, {\it et al.}, Phys. Rev. Lett. {\bf 70},
                   1553 (1993).
\bibitem{Wollman}  D.A. Wollman, {\it et al.}, Phys.
                   Rev. Lett. {\bf 71}, 2134 (1993).
\bibitem{Eliashberg} G.M. \'Eliashberg, Zh. Eksp. Teor. Fiz.
		   {\bf 38}, 966 (1960) [Sov. Phys. JETP {\bf 11}, 696
		   (1960)].
\bibitem{SP}	   N.E. Bickers, D.J. Scalapino and S.R. White,
		   Phys. Rev. Lett. {\bf 62}, 961 (1989). P. Monthoux
and  D.  Pines, {\it ibid.} {\bf 69}, 961 (1992).
\bibitem{anyon}	   D.S. Rokhsar, Phys. Rev. Lett. {\bf
		   70}, 493 (1993).
\bibitem{Hardy}    W.N. Hardy, {\it et al.}, Phys.
                   Rev.  Lett. {\bf 70}, 3999 (1993).
\bibitem{Yigal1}   Y. Meir and N.S. Wingreen, Phys. Rev. Lett. {\bf 68},
		   2512 (1992).
\bibitem{Andreev}  A.F. Andre\'{e}v, Zh. Eksp. Teor. {\bf 46}, 1823 (1964)
                   [Sov. Phys. JETP {\bf 19}, 1228 (1965)].
\bibitem{othAR} Recent theoretical
[F.W.J. Hekking {\it et al.}, Phys. Rev. Lett. {\bf 70}, 4138 (1993)]
and experimental
[J.M. Hergenrother,
M.T. Tuominen and M. Tinkham, Phys. Rev. Lett. {\bf 72}, 1742 (1994)]
investigations of Andre\'ev scattering in
normal-superconductor-normal systems have focussed on
small islands of low-temperature superconductor.
Coulomb blockade plays a large role in the response of these
systems and gap anisotropy plays no role.
\bibitem{Aproc} e.g. Y. De Wilde, {\it et al.}, Phys. Rev. Lett.
{\bf 72}, 2278 (1994).
\bibitem{Klein} From microwave measurements the mean-free path is $1.6\mu$
[D.A. Bonn, {\it et al.}, Phys. Rev. B {\bf 47}, 11314
(1993)]. Measurements of similar quality on ${\rm
Bi_2Sr_2CaCu_2O_8}$ have not been done, although regions of
the surface without imperfections $500$\AA$\times 500$\AA\
are readily found with STM [C.K. Shih, private
communication].
\bibitem{Niu} Q. Niu, M.C. Chang, and C.K. Shih, Phys. Rev. B to
be published.
\end{references}
\end{document}